\documentclass[aps,pra,showpacs,twocolumn]{revtex4-1}

\usepackage{dsfont}
\usepackage{hyperref}
\usepackage{amssymb,amsmath,amsfonts}
\usepackage{mathrsfs}
\usepackage[none]{hyphenat}

\usepackage{color}

\usepackage[normalem]{ulem}

\usepackage[caption=false]{subfig}
\newcommand{\id}{\mathds{1}}

\DeclareMathOperator{\Tr}{Tr}

\newcommand{\ket}[1]{\left|{#1}\right\rangle}
\newcommand{\bra}[1]{\left\langle{#1}\right|}

\newcommand{\oost}{\frac{1}{\sqrt{2}}}

\newcommand{\hd}{\ensuremath{\mathsf{h}}}
\newcommand{\xd}{\ensuremath{\mathsf{x}}}
\newcommand{\yd}{\ensuremath{\mathsf{y}}}
\newcommand{\zd}{\ensuremath{\mathsf{z}}}

\newcommand{\matel}[2]{\left|#1\middle\rangle\middle\langle#2\right|}

\begin{document}
%
\title{
       Exponentially many entanglement and correlation constraints for 
       multipartite quantum states
         }

\author{
        Christopher Eltschka$^1$, Felix Huber$^{2,3,4}$, Otfried G\"uhne$^4$,
        and Jens Siewert$^{5,6}$  
       }
\affiliation{
$^{1}$ Institut f\"ur Theoretische Physik, Universit\"at Regensburg, D-93040 Regensburg, Germany\\
$^{2}$ ICFO - Institut de Ci\`encies Fot\`oniques, The Barcelona 
       Institute of Science and Technology, 
       E-08860 Castelldefels (Barcelona), Spain\\
$^{3}$ Institut f\"ur Theoretische Physik, Universit\"at zu K\"oln,
       D-50937 K\"oln, Germany\\
$^{4}$ Naturwissenschaftlich-Technische Fakult\"at, Universit\"at Siegen, 
D-57068 Siegen, Germany\\
$^{5}$ Departamento de Qu\'{i}mica F\'{i}sica, Universidad del Pa\'{i}s Vasco UPV/EHU, E-48080 Bilbao, Spain\\
$^{6}$ IKERBASQUE Basque Foundation for Science, E-48013 Bilbao, Spain
}

\date{\today}
\begin{abstract}
We present a family of correlations constraints that apply to all 
multipartite quantum systems of finite dimension. 
The size of this family is exponential
in the number of subsystems.
We obtain these relations by defining and
investigating the generalized state inversion map. 
This map provides a systematic way to generate 
local unitary invariants of degree two in the state and 
is directly linked to the shadow inequalities
proved by Rains~[IEEE Trans.\ Inf.\ Theory {\bf 46}, 54 (2000)].
The constraints are stated in terms of linear inequalities for the linear
entropies of the subsystems. For pure quantum states  they turn 
into monogamy relations that constrain the distribution of
bipartite entanglement among the subsystems of the global state.
\end{abstract}

\pacs{
     }

\maketitle

\emph{Introduction}.---%
%
The discussion of entanglement monogamy 
started more than two decades ago~\cite{Bennett1996}.
Its first precise quantitative formulation was given by
Coffman et al.\  in an {\em equality} for the distribution
of entanglement among three qubits~\cite{CKW2000}, whereas its weaker form,
an inequality, subsequently was generalized by Osborne
and Verstraete~\cite{Osborne2006} to an arbitrary number of qubits.
In the meantime, there have been numerous attempts to
extend the results of Refs.~\cite{CKW2000,Osborne2006} or
to find new independent monogamy relations,
see, e.g., 
Refs.~\cite{KW2004,Ou2007,Gour2007,Kim2009,Eltschka2009,Sanders2010,Bai2014,Fei2015,ES2015,Adesso2016,Lancien2016,Luo2016,Fei2017,Meyer2017,Gour2017,Huber2017,ES2018}. 
Furthermore, it was found that
also correlations other than entanglement, such as nonlocality,
may obey monogamy relations~\cite{Toner2009,Seevinck2010,Cabello2017}.

Some authors consider monogamy of correlations an inherent
property of quantum mechanics~\cite{Ujjwal2016}, 
however, there are results that
seem to challenge this point of view:
(a) The fundamental monogamy relations by Coffman et al.~\cite{CKW2000} 
and Osborne and Verstraete~\cite{Osborne2006} cannot 
straightforwardly be generalized
to local dimensions higher than two~\cite{Ou2007}, (b) faithfulness of
entanglement measures and monogamy properties seem to be 
mutually exclusive~\cite{Lancien2016}, and (c)
systematically including contributions
of multipartite entanglement appears to be 
difficult~\cite{Eltschka2009,Adesso2016}.
This raises also the question of what the general form of a monogamy relation
should be~\cite{Eltschka2009,Bai2014,Lancien2016,Ujjwal2016,Gour2017}. 
Generally, it is assumed that the terms characterizing different 
correlations have to be 
added, possibly after raising each term to some fixed power. 
Again, this seems to contradict the recent finding that general monogamy 
equalities, as well as inequalities, for any number of 
qubits~\cite{ES2015} and even 
higher-dimensional systems~\cite{ES2018} exist whose terms are summed
with alternating signs.

In the present work, we adopt the viewpoint that {\em any} functional 
relation between quantifiers for different correlations (equality
or inequality) may be considered a monogamy relation, 
simply because it constrains the free distribution of these 
correlations among the parties of a multipartite system.
The relevant point is that the terms in the relation 
are of physical significance. 
If, for example, all the terms are related to measures
of entanglement in different subsets of the parties, 
one would call the correlation constraint a monogamy 
relation for entanglement, because it describes restrictions regarding
the distribution of entanglement among the parties. 
An illuminating example that this approach is sensible is that certain 
correlation constraints in 
Refs.~\cite{ES2015,ES2018} hold both for pure and mixed states,
but represent monogamy relations for entanglement only in the case
of pure states.

It is natural to expect a variety of correlation
constraints 
originating from the algebraic properties of the 
density matrix, such as the positivity of the state. 
Based on this intuition, our objective is to devise a method to systematically
generate an entire family of correlation constraints.
Our central results show that the 
positivity condition under certain mappings alone gives rise to an exponential 
number of independent correlation constraints, as well as to
monogamy relations for entanglement in multipartite pure states of 
any number of parties and finite local dimension. Our method to derive these 
relations is based on and extends the so-called universal state 
inversion~\cite{Horodecki1999,Rungta2001}. 
It turns out that the generalized inverter map is directly
related to Rains' 
shadow inequalities~\cite{Rains1998,Rains2000} which, by virtue
of these investigations, can be assigned a direct physical interpretation.

We introduce universal state inversion by explaining relevant examples for
entropy inequalities that can 
directly be derived from the inverted state. 
It is then straightforward to understand
the definition and properties of the generalized state 
inversion map.
After presenting our main results we discuss
several routes of investigation that get new input through
our findings; these include detection of entanglement, 
derivation of new inequalities for the linear entropy, and the
quantum marginal problem.

{\em Entropy inequalities from universal state inversion.}---%
In the following we consider normalized
states of an $N$-partite system 
$\rho\in \mathcal{B}(\mathcal{H}_1\otimes\ldots\otimes\mathcal{H}_N)$,
where $\mathcal{H}_j$ are Hilbert spaces with
$\dim\mathcal{H}_j=d_j$ ($j=1,\ldots,N$), and $\Tr(\rho)=1$.
Let us start with bipartite systems, 
$N=2$. We denote the global state by $\rho_{12}$ while the reduced state of
the first subsystem is $\rho_1=\Tr_2\left(\rho_{12}\right)$, and
analogously $\rho_2=\Tr_1\left(\rho_{12}\right)$. In Ref.~\cite{Rungta2001}
it was shown that the operator
\begin{equation}
   \tilde{\rho}_{12}\ = \id_{12}-\rho_1\otimes\id_2-\id_1\otimes\rho_2
                        +\rho_{12}\ \geqq\ 0\ \ ,
\label{eq:twoparties}
\end{equation}
is positive semidefinite. Here, $\id_j$ is the identity operator acting
on subsystem $j=1,2$, and $\id_{12}$ the one for the full system. 
By multiplying Eq.~\eqref{eq:twoparties} by $\rho_{12}$ and applying the
trace as well as the definition for the linear entropy of subsystem $j$,
$\tau_j=2\left[1-\Tr\left(\rho_j^2\right)\right]$, one obtains the well-known
subadditivity of linear entropy~\cite{Audenaert2007}
\begin{align}
    \tau_{12}\ \leqq\ \tau_1\ +\ \tau_2\ \ .
\label{eq:subadd}
\end{align}
One recognizes the usefulness of the operator $\tilde{\rho}_{12}$,
the result of the {\em universal state inversion map} 
applied to the state $\rho_{12}$.
It arises through subsequently tracing out all of the subsets of parties
and padding with identities, 
multiplying by $(-1)$
per trace operation and adding up the results. Analogously, universal state
inversion for a three-party state $\rho_{123}$ yields (for the sake of brevity
we drop the tensor factors $\id_j$ and identify $\id_{123}\equiv\id$)
\begin{equation}
   \tilde{\rho}_{123} = \id-\rho_1-\rho_2-\rho_3
                        +\rho_{12}+\rho_{13}+\rho_{23}-\rho_{123}
                          \geqq 0\ .
\label{eq:threeparties}
\end{equation}
In analogy with the operations above the inequality 
\begin{align}
\tau_1+\tau_2+\tau_3+\tau_{123}\ \geqq\ \tau_{12}+\tau_{13}+\tau_{23}
\label{eq:stronglin}
\end{align}
is found~\cite{ES2018}.
It resembles a symmetrized and reversed version of the 
strong subadditivity 
inequality for the von Neumann entropy $S$~\cite{Lieb1973}, 
which reads $S_{123}+S_2 \leqq S_{12}+S_{23}$. We mention that
the analogue of inequality~\eqref{eq:stronglin} 
for von Neumann entropy was discussed 
by Hayden et al.~\cite{Hayden2013}, as a quantum extension 
to the so-called 
interaction information~\cite{McGill1954}
and a desirable monogamy property 
in the context of holographic theories.

Summarizing this introduction, one can use the positivity of
the universal state inversion 
map~\cite{Rungta2001,Hall2005,Hall2006,ES2018}
\begin{align}
   \mathcal{I}(\rho)\ =\ \Big[  \prod_{j=1}^N
                        \big( \Tr_j(\cdot)\otimes\id_j 
                              - \mathrm{id} 
                                     \big)
                         \Big] \ \rho
\label{eq:rhotildeprod}
\end{align}
to derive relevant inequalities, or {\em correlation constraints},
for arbitrary states of multipartite systems of any finite local
dimension. In Eq.~\eqref{eq:rhotildeprod}, $\mathrm{id}$ denotes the
identity map. An alternaltive way of writing 
the map $\mathcal{I}(\rho)$ 
in terms of reduced states $\rho_S=\Tr_{S^c}\left(\rho\right)$ is
\begin{align}
     \mathcal{I}(\rho)\ =\ \sum_{S\subseteq\{1\ldots N\}}\ (-1)^{|S|}\ 
                            \rho_S
                                            \otimes 
                            \id_{S^c}
\ \ ,
\label{eq:rhotildered}
\end{align}
where $S$ is a set of subsystem indices, $S^c$ is its complement 
$S^c=\{1\ldots N\}\setminus S$,
and $|S|$ denotes the cardinality of $S$. From Eq.~\eqref{eq:rhotildered}
it is evident that $\mathcal{I}(\rho)$ commutes with local unitary 
operations~\cite{Rungta2001,Hall2005,ES2018}. 
In what follows we will generalize the inversion map and obtain 
a powerful tool for the analysis of correlations in arbitrary
finite-dimensional multi-party states.
%
%
%

{\em Generalized $T$-inverter.---}%
We obtain a more general form of
the state inversion map, Eq.~\eqref{eq:rhotildeprod}, by
reversing the minus sign in some of the factors.
Assume we retain a minus sign only for all those subsystem indices
that are contained in $T\subseteq \{1\ldots N\}$, the other factors
come with a plus sign.
Then the generalized {\em T-inversion map} 
$\mathcal{I}_T(\cdot)$ can elegantly be written as
\begin{align}
   \mathcal{I}_T(\rho)\ =\ & \Big[ \prod_{j=1}^N 
                             \Big(
                           \Tr_j(\cdot)\otimes \id_j
                           +(-1)^{|T\cap \{j\}|}\mathrm{id} 
                             \Big)
                             \Big]\ \rho
\label{eq:T-inverterA}\\
                      \ =\ & \sum_{S\subseteq\{1\ldots N\}}
                           (-1)^{|S\cap T|} 
                           \left(\Tr_{S^c}\rho\right)\otimes \id_{S^c}
\ \ . 
\label{eq:T-inverterB}
\end{align}
The original state inverter Eq.~\eqref{eq:rhotildeprod}
is found for $T=\{1\ldots N\}$. 

Interestingly, there exists a representation of the map $\mathcal{I}_T(\cdot)$
in Kraus form, which we will derive now. First, let us consider the
representation of a single factor in Eq.~\eqref{eq:T-inverterA} acting
on a $d$-level system.
To this end, we note that for a complete basis of traceless 
Hermitian matrices,
$\{\hd_m\}$ complemented by $\hd_0\equiv \id$, with
$\Tr\left(\hd_m\hd_n\right)=d\delta_{mn}$,
and a Hermitian operator $A$ 
we have~\cite{ES2018}
\begin{align}
  \label{eq:tracegens}
  \Tr (A)\id &\ =\  \frac{1}{d}\sum_{m=0}^{d^2-1} 
             \mathsf{h}_m\ A\ \mathsf{h}_m\ \ ,\\
  \label{eq:transgens}
  A^T &\ =\  \frac{1}{d}\sum_{m=0}^{d^2-1} 
             \mathsf{h}_m^T\ A\ \mathsf{h}_m
\ \ .
\end{align}
With these relations it is easy to find the action of the 
$j$th factor in Eq.~\eqref{eq:T-inverterA} on $A_j$ 
(a Hermitian operator that acts on a $d_j$-dimensional Hilbert space),
\begin{align}
\label{eq:chan1}
    \Tr(A_j)\id_j -A_j\ =\ & 
               \frac{2}{d_j}   \sum_{k<l}^{d_j-1} \yd_{kl}\ A_j^*\ \yd_{kl}
\\
\label{eq:chan2}
    \Tr(A_j)\id_j +A_j\ =\ & 
               \frac{2}{d_j} \Big[\  A_j^*\ +\ \sum_{k<l}^{d_j-1} \xd_{kl}\ 
      A_j^*\ \xd_{kl}\ +   \Big.
\nonumber\\
       & +\ 
                \Big.\ \sum_{k=1}^{d_j-1} \zd_{k}\ A_j^*\ \zd_{k}\ \Big]
\ \ ,
\end{align}
where we have used the generalized Gell-Mann matrices~\cite{GellMann}. 
%
%

In Eqs.~\eqref{eq:chan1},~\eqref{eq:chan2} we clearly 
observe the Kraus form 
of the map. Note that it is applied to the complex conjugate $A_j^*$.
Since all the factors in Eq.~\eqref{eq:T-inverterA} commute, the
Kraus form extends to the entire operator product, with the Kraus
operators on the full system being tensor products of the single-system
generators (for details see Appendix A~\cite{supplement}). Thus, the map 
$\mathcal{I}_T(\cdot)$ on the full system can be written as
$\mathcal{I}_T=\Lambda\circ K$ where $\Lambda$ is the Kraus map
and $K$ is the complex conjugation. The existence of this representation
proves the positivity of the generalized $T$-inversion map, 
Eqs.~\eqref{eq:T-inverterA} and \eqref{eq:T-inverterB}.
We mention also that, on application of this map to Hermitian operators,
the complex conjugation may be replaced by a transposition (see also below). 
In this sense, first transposing the state and subsequently
applying $\mathcal{I}_T(\cdot)$ may be viewed as
a generalization of the Werner-Holevo channel~\cite{AHW2000} 
to multipartite systems,
and is a completely positive map.

A consequence of the positivity of $\mathcal{I}_T(\cdot)$ is that,
for two semidefinite positive operators $M_1$, $M_2$, 
we have~\cite{footnote-shadow}
\begin{align}
            \Tr\big[\ M_1\ \mathcal{I}_T(M_2)\ \big] \ \geqq\ 0
\ \ .
\label{eq:positive}
\end{align}
By inserting Eq.~\eqref{eq:T-inverterB} and noting that
$\Tr\left[M_1\Tr_{S^c}\left(M_2\right)\right]=
\Tr\left[\Tr_{S^c}\left(M_1\right)\Tr_{S^c}\left(M_2\right)\right]$
we obtain
\begin{align}
   \sum_{S\subseteq\{1\ldots N\}}
                           (-1)^{|S\cap T|} 
        \Tr\big[ \Tr_{S^c}\! \left(M_1\right) \Tr_{S^c} 
                          \! \left(M_2\right) \big] \geqq 0
\ . 
\label{eq:shadow}
\end{align}
That is, as a by-product of the definition of generalized $T$-inversion
we have derived Rains's shadow inequalities,
Eq.~\eqref{eq:shadow}~\cite{Rains1998,Rains2000},
which are an important tool for investigating the existence of quantum
error correcting 
codes~\cite{Sloane1990,Dougherty1995,Rains1998b,Rains1999,Huber2018}.
It is remarkable that the shadow inequalities are directly 
linked to generalized state 
inversion, 
which in turn is connected with 
correlation and entanglement distribution constraints, as we will show below.
The shadow inequalities provide a quick alternative proof
for the positivity of the generalized $T$-inversion map:
Choose $M_1=\ket{\psi}\!\bra{\psi}$, $M_2=\rho$ in Eq.~\eqref{eq:shadow}, 
where $\ket{\psi}$ is an arbitrary finite-dimensional pure state, 
and $\rho$ is an arbitrary state of the same multi-party system.
This yields $\bra{\psi} \mathcal{I}_T(\rho)\ket{\psi}\geqq 0$,
implying positivity of $\mathcal{I}_T(\rho)$.

There is another interesting property of the generalized 
$T$-inversion, which may be called `coarse graining'. Consider,
for example, the tripartite state $\rho_{123}$ for which we may combine
(coarse grain) the subpartitions \mbox{$2$ and $3$} into a single partition, so
that we end up with a bipartite state $\rho_{1(23)}$. Suppose we want
to apply an inversion map with $T^{(\mathrm{coarse})}=\{(23)\}$ 
to the coarse-grained state $\rho_{1(23)}$ [i.e.,
$\mathcal{I}_{\{(23)\}}(\rho_{1(23)})]$, can we build it
from the inverted states on the fine-grained system? The answer is positive: 
One has to average over all those $T$-inverted states of the 
fine-grained system with the following rules
for each set of subsystem indices grouped into a single party: 
(a)~the sets $T$ characterizing the fine-grained inversions
have odd parity for those single-system indices 
appearing in $T^{(\mathrm{coarse})}$;
(b)~for the single-system indices not appearing in $T^{(\mathrm{coarse})}$
the parity in the fine-grained $T$ sets has to be even.
Hence, in our example $\mathcal{I}_{\{(23)\}}(\rho_{1(23)})
        =\frac{1}{2}\left[\mathcal{I}_{\{2\}}(\rho_{123})+
                     \mathcal{I}_{\{3\}}(\rho_{123})\right]$.
As special cases of this property we have for an $N$-partite
state $\rho$
\begin{subequations}
\begin{align}
         \id\ -\ \rho\ = & \ \frac{1}{2^{N-1}}
        \sum_{T\subseteq\{1\ldots N\},\ |T|\  \small \mathrm{odd}
             }
                             \mathcal{I}_T(\rho)
\\
         \id\ +\ \rho\ = & \ \frac{1}{2^{N-1}}
        \sum_{T\subseteq\{1\ldots N\},\ |T|\  \small \mathrm{even}
             }
                             \mathcal{I}_T(\rho)
\ \ .
\end{align}
\end{subequations}
We present a detailed proof in Appendix~B~\cite{supplement}.

{\em Exponentially many correlation constraints.---}%
Consider the special case
$M_1=M_2=\rho$ of Eq.~\eqref{eq:positive},
\begin{equation}
\Tr\left[ \rho\ \mathcal{I}_T(\rho)\right]\geqq 0
\ \ .
\label{eq:poscon}
\end{equation}
This observation 
gives rise to a notable set of constraints on the possible
correlations in a multipartite state. 
We use the decomposition of the generalized inverted state 
in Eq.~\eqref{eq:T-inverterB} 
as well as the definition of the linear entropy
to expand 
Eq.~\eqref{eq:poscon} 
and find (for $T\neq \emptyset$)
\begin{align}
 0\ \leqq\  & \sum_{S\subseteq\{1\ldots N\}}
                           (-1)^{|S\cap T|} 
                \Tr\left[ \rho \Tr_{S^c}\left(\rho\right)
                   \right]
\nonumber\\
                      \ =\ & 
                           \sum_{S\subseteq\{1\ldots N\}}
                           (-1)^{|S\cap T|} 
                \Tr\left( \rho_S^2 \right)
\nonumber\\  
                      \ =\ &
             \frac{1}{2}\sum_{\emptyset\neq S\subseteq\{1\ldots N\}}
                             (-1)^{|S\cap T|+1}\ \tau_S
\ \ .
\label{eq:newcorr}
\end{align}
For each choice of $T\neq\emptyset$ , this is a constraint for the correlations
across the bipartite splits $S|S^c$ as quantified by the linear entropy, 
with different distribution of minus signs. 
Altogether these are $2^N-1$ relations
(the condition for $T=\emptyset$ is trivial in view of the fact that all
subsystem purities are positive). We show in Appendix~C that the
right-hand sides of these inequalities are functionally independent.

The exponentially many  correlation constraints in 
Eq.~\eqref{eq:newcorr} constitute the first key result of our work. 
These are necessary conditions related to the quadratic local unitary invariants
$\Tr\left(\rho_S^2\right)$ of any finite-dimensional multi-party quantum state.
It is particularly satisfactory that these conditions 
do not originate from ad hoc 
assumptions regarding their functional form. 
Rather, they arise systematically
through the definition and algebraic properties of generalized $T$-inversion.


%
{\em Local unitary invariants $\mathscr{C}_T(\rho)$.---}%
%
The generalized $T$-inverter
commutes with local unitaries, so that 
we can define the  local unitary invariants
\begin{align}
 \mathscr{C}_T(\rho)\ \equiv\ \sqrt{\Tr\big[\rho\ \mathcal{I}_T(\rho) 
                                \big]}
\ \ .
\label{eq:CT}
\end{align}
%

%
As we have seen, these invariants are relevant because they
generate the correlation constraints
Eq.~\eqref{eq:newcorr}. Therefore, let us 
briefly mention some properties of $\mathscr{C}_T(\rho)$.

A direct consequence of the Kraus form of $\mathcal{I}_T(\cdot)$
is that, for pure states
$\rho_{\psi}=\ket{\psi}\!\bra{\psi}$, the local invariant $\mathscr{C}_T(\psi)$
vanishes whenever there is an odd number of factors with a minus sign
in Eq.~\eqref{eq:T-inverterA},
\begin{align}
\mathscr{C}_T(\psi)\ =\ 0\ \ \ \mathrm{for}\ \ \ |T|\ \equiv\ 1\ \pmod{2}
\ \ 
\label{eq:zero}
\end{align}                               
(we give a proof of this fact in Appendix D~\cite{supplement}).

Furthermore, we recall that in Ref.~\cite{ES2018} the
distributed concurrence 
$C_D(\psi)\equiv \sqrt{\Tr\left[\rho_{\psi}\ \mathcal{I}_{\{1\cdots N\}}(\rho_{\psi})\right]}$
was defined, which generalizes the well-known (bipartite) 
concurrence~\cite{Rungta2001}
and in certain cases is an entanglement monotone, that is,
non-increasing on average under stochastic local operations and
communication.
Because of the apparent analogies with Eq.~\eqref{eq:CT}
the question arises 
whether the other invariants $\mathscr{C}_T(\psi)$ possibly are entanglement 
monotones.
The answer is: None of the local invariants $\mathscr{C}_T(\psi)$
with $T\neq \{1 \ldots N\}$ can be an entanglement monotone.
The proof will also be shown in Appendix D~\cite{supplement} 
and essentially relies on the factorization property
\begin{align}
     \mathscr{C}_T(\rho_{\mathrm{prod}})\ =\ 
     \mathscr{C}_{T_S}(\rho_S) \mathscr{C}_{T_{S^c}}(\rho_{S^c})
\end{align}
for product states $\rho_{\mathrm{prod}}=\rho_{S}\otimes\rho_{S^c}$
{\em Monogamy of entanglement.---}%
%
Even though $\mathscr{C}_T(\psi)$ is not in general
an entanglement monotone,
Eq.~\eqref{eq:newcorr} leads to exponentially many monogamy relations
for the entanglement in the pure state $\ket{\psi}$. This is because
for a pure state the linear entropy of subsystem $S$ equals the
squared concurrence over the bipartite split $S|S^c$, that is,
$\tau_S=2[1-\Tr\left(\rho_S^2\right)]\equiv C_{S|S^c}^2$.

Thus we have the second main result of our article,
the $2^N-1$ monogamy inequalities
\begin{align}
 0\ \leqq\ \sum_{\emptyset\neq S\subset\{1\ldots N\}}
                             (-1)^{|S\cap T|+1}\ C_{S|S^c}^2(\psi)
\ \ ,
\label{eq:newmonog}
\end{align}
one inequality for each $\emptyset\neq T\subseteq \{1\ldots N\}$,
which are valid for any number of parties $N$ 
and any finite local dimensions. These inequalities constrain the
distribution of concurrence among the subsystems of any global
pure state $\ket{\psi}$. They are related to the local unitary invariants of
homogeneous degree two in the state. Note, however, that according
to their definitions the invariants $\mathscr{C}_T(\psi)$ and 
the concurrence $C_{S|S^c}(\psi)$ are of homogeneous degree one in
the state $\rho_{\psi}$, whereas the relations~\eqref{eq:newmonog}
[as well as Eq.~\eqref{eq:newcorr}] are of homogeneous
degree two, just as the results, e.g., in
Refs.~\cite{CKW2000,Osborne2006,ES2015,ES2018}.
Again, there is no ad hoc assumption underlying these
constraints, they naturally follow from the algebraic properties
of the generalized $T$-inverter.

{\em Entanglement detection.---}%
%
In Ref.~\cite{Horodecki1999} it was first observed that the reduction
map $\Lambda(\rho)=\id-\rho$ is positive, but not completely positive:
For non-separable bipartite states 
$\rho_{AB}\in \mathcal{B}(\mathcal{H}_A\otimes\mathcal{H}_B)$
we may have $(\id\otimes\Lambda)(\rho_{AB})\ngeqq 0$. 
Consequently, this map can be used to detect entanglement in the 
state $\rho_{AB}$ (reduction criterion). 
Later, similar maps were 
studied~\cite{Hall2006,Breuer2006,Wolf2016,Lewenstein2016,Huber2016}.
As we have seen above, the state inversion map can be represented
by a concatenation of a transposition and a subsequent Kraus channel.
This elucidates that also the reduction criterion includes 
a partial transposition. 
While the reduction criterion is well known to detect
fewer entangled states than the positive partial 
transpose criterion, the states it detects are guaranteed
to be distillable~\cite{Horodecki1999}.
In the spirit of~Ref.~\cite{Lewenstein2016} we may even further
generalize the $T$-inversion map by introducing real numbers
$0\leqq \alpha_j,\beta_k\leqq 1$, 
\begin{align}
   \mathcal{I}_T^{\{\alpha_j,\beta_k\}} =
           \prod_{j\in T} \bigg[ \Tr_j(\cdot)\id_j-\alpha_j\mathrm{id}\bigg]
           \prod_{k\notin T} \bigg[ \Tr_k(\cdot)\id_k+\beta_k\mathrm{id}\bigg]
\end{align}
which again is a positive, but not completely positive map.
The strength of the corresponding entanglement criteria 
(in analogy with the bipartite case investigated in Ref.~\cite{Lewenstein2016})
and the question how to choose the optimal $\alpha_j$, $\beta_k$ will 
be discussed in future work.

{\em More entropy inequalities.---}
Based on the positivity of generalized
$T$-inversion, more relevant inequalities for the linear entropy can be derived.
Note that the linear entropy is proportional to
the Tsallis 2-entropy~\cite{Tsallis1988,Audenaert2007} 
and has a simple functional relation with
the R\'enyi $\alpha$-entropy for $\alpha=2$~\cite{Wehrl1978}. 
To date, only few inequalities are known for the
linear entropy~\cite{Audenaert2007,Gour2007,Petz2015,Luo2016,Huber2017}.

Recall that for a bipartite state $\rho_{12}$ the $\{12\}$-inverter
leads to Eq.~\eqref{eq:subadd}. Analogously, we may use the positivity
of the $\{1\}$-inverter as well as the $\{2\}$-inverter,
and find
%
     $\tau_{12} \geqq \left|\tau_1 - \tau_2\right|$,
%
leading to the linear entropy analogue of the Araki-Lieb triangle 
inequality~\cite{Lieb1970} found by Audenaert~\cite{Audenaert2007}
\begin{align}
     \left|\tau_1 - \tau_2\right|\ \leqq\
     \tau_{12} \ \leqq\
     \tau_1\ +\ \tau_2 \ \ .
\end{align}

Now let us go back to three-party states $\rho_{123}$, for which we
have found Eq.~\eqref{eq:stronglin}.
We note that the linear entropy analogue of strong subadditivity, 
$\tau_2+\tau_{123}\leqq\tau_{12}+\tau_{23}$, 
does {\em not} hold~\cite{Petz2015}; 
this is readily demonstrated by analyzing the three-qubit state 
$\rho_{\mathrm{II}}=
     \ket{\Phi^+_{12}}\!\bra{\Phi^+_{12}}\otimes\frac{1}{2}\id_3$,
with $\ket{\Phi^+}=\oost\left(\ket{00}+\ket{11}\right)$. The reverse
inequality does not hold either, as the state 
$\rho_{\mathrm{III}}=
     \ket{\Phi^+_{13}}\!\bra{\Phi^+_{13}}\otimes\frac{1}{2}\id_2$
shows.
Yet, 
%
%
we can add the relations for the coarse-grained
$\mathscr{C}^2_{\{(12)\}}(\rho_{(12)3})$ and
$\mathscr{C}^2_{\{(23)\}}(\rho_{1(23)})$, where partitions $(12)$ and $(23)$,
respectively, are considered a single party, 
and obtain
\begin{align}
\tau_1 + \tau_3\ \leqq\ \tau_{12}+\tau_{23}+2\tau_{123}\ \ .
\label{eq:linUhl}
\end{align}
This relation is reminiscent of the  weak monotonicity
$S_1+S_3\leqq S_{12}+S_{23}$ for von Neumann entropies, 
which is equivalent to strong subadditivity~\cite{Lieb1973,Uhlmann1973}.
Alternatively, we
can purify the state $\rho_{123}$ with a fourth party, use
$\tau_{1234}=0$, $\tau_{123}=\tau_4$ and $\tau_{23}=\tau_{14}$,
and then re-label the parties $1\leftrightarrow 2$, 
                              $3\leftrightarrow 4$, so that
\begin{align}
\tau_{2}+\tau_{123}\ \leqq\ \tau_{12}+\tau_{23}+2\tau_3
\ \ .
\end{align}
The latter result shows the correction in a linear entropy inequality 
analogous to the standard strong subadditivity relation for the 
von Neumann entropy.
%

{\em Compatibility of marginals.---}%
Finally we want to highlight the relation of our results with
the quantum-marginal problem, that is, the question whether or not
a given set of reduced states is compatible with a joint global
state~\cite{Klyachko2006}. Clearly, our 
linear-entropy constraints 
\eqref{eq:newcorr} represent necessary conditions for the reduced
states $\rho_S$ to be compatible with the global state $\rho$.
However, we can make new statements even at the operator level.

In order to see this, consider again a three-party state $\rho_{123}$. 
Butterley et al.~\cite{Butterley2006} found for three qubits that, 
given a set of 
two-body marginals $\rho_{12}$, $\rho_{23}$ and $\rho_{13}$, compatibility
with a joint state $\rho_{123}$ requires positivity of the operator
\begin{equation}
   \Delta = \id-\rho_1-\rho_2-\rho_3
                        +\rho_{12}+\rho_{13}+\rho_{23}
                        \  \geqq\ 0\ .
\label{eq:butterley1}
\end{equation}
By comparing with Eq.~\eqref{eq:threeparties} we note that this
follows immediately for arbitrary local dimensions 
from the positivity of 
$\tilde{\rho}_{123}\equiv\mathcal{I}_{\{123\}}(\rho_{123})$,
and hence 
\mbox{$\mathcal{I}_{\{123\}}(\rho_{123})+\rho_{123}\geqq 0$.} 
Now, invoking generalized $T$-inversion for odd integer $|T|$, e.g.,
$\mathcal{I}_{\{1\}}(\rho_{123})+\rho_{123}\geqq 0$ gives
\begin{equation}
   \id-\rho_1+\rho_2+\rho_3
                        -\rho_{12}-\rho_{13}+\rho_{23}
                        \  \geqq\ 0\ ,
\label{eq:butterley2}
\end{equation}
and analogous relations for 
$\mathcal{I}_{\{2\}}(\rho_{123})$ and $\mathcal{I}_{\{3\}}(\rho_{123})$.
This construction generalizes to an arbitrary number of parties:
From $\mathcal{I}_T(\rho) + \rho \geqq 0$ for odd $|T|$, one obtains 
exponentially many independent operator
constraints for the compatibility of quantum marginals~\cite{Huber2017Thesis}.

{\em Conclusions.---}%
We have extended the theory of universal quantum state inversion
by defining the generalized $T$-inversion map $\mathcal{I}_T$.
This map turns out to be a unifying building block for various aspects of
quantum correlations in finite-dimensional multi-party systems:
It brings together the theory of multipartite entanglement and
entanglement monogamy with a formalism originating from 
quantum error correcting codes, and also the quantum marginal problem.
Thereby it elucidates the common algebraic origin of the different physical
properties investigated in these fields. Most prominently, it provides
a systematic way to generate and explore correlation constraints
and monogamy relations for entanglement in composite systems of 
arbitrary finite local dimension. We mention the immediate
application of the constraints in Eq.~\eqref{eq:newmonog} in
excluding the existence of absolutely maximally entangled states
for certain party numbers  and local dimensions~\cite{Huber2018}.

\acknowledgments
This work was funded by the German Research Foundation Project
EL710/2-1 (C.E., J.S.), by Swiss National Science Foundation 
(Doc.Mobility Grant No. 16502) (F.H.),
ERC Consolidator Grant 683107/TempoQ (O.G.), and
by Basque Government grant IT986-16,
MINECO/FEDER/UE grant  FIS2015-67161-P,
and UPV/EHU program UFI 11/55 (J.S.). 
The authors would like to thank Marcus Huber and Nikolai Wyderka
for stimulating discussions. C.E.\ and J.S.\ acknowledge
Klaus Richter's support for this project. 
%
%
%
%
%
%
%
%
\renewcommand{\thesection}{\Alph{section}}
\renewcommand{\theequation}{\thesection\arabic{equation}}
\appendix
\section{Kraus form of the generalized $T$-inversion map}
%
The procedure to derive the Kraus representation of the 
generalized $T$-inversion map is analogous to that for
the universal state inverter, cf.~Ref.~\cite{ES2018}.
We use the definitions of the generalized Gell-Mann matrices
in $d$ dimensions
%
\begin{align*}
  \mathsf{x}_{kl} = &
  \sqrt{\frac{d}{2}}(\matel{k}{l}+\matel{l}{k})\ \ ,
\\
  \mathsf{y}_{kl} = &
  \sqrt{\frac{d}{2}}(-\mathrm i\matel{k}{l}+\mathrm i\matel{l}{k})\ \ ,
\\
 \mathsf{z}_l  =  &\sqrt{\frac{d}{l(l+1)}}
  \left(-l\matel{l}{l} + \sum_{k=0}^{l-1}\matel{k}{k}\right) \ \ ,
\end{align*}
where $(0\leqq k < l < d)$. 
Now we re-label them as follows:
\begin{align}
  \label{Seq:generators}
  \mathsf{h}_0 &\ = \ \id\ \ ,\\
  \mathsf{h}_{l^2+2k} &\ = \ \mathsf{x}_{kl}\ \ ,\\
  \mathsf{h}_{l^2+2k+1} &\ =\  \mathsf{y}_{kl}\ \ ,\\
  \mathsf{h}_{l^2+2l} &\ =\  \mathsf{z}_l\ \ .
\end{align}
Hence, the expressions for the factors of the inverter,
Eqs.~(11), (12) in the main text, take the form
\begin{align}
\label{Seq:chan1}
    \Tr(A_j)\id_j -A_j\ =\ & 
               \frac{2}{d_j}   \sum_{k=0}^{d_j-2}\sum_{l=k+1}^{d_j-1} 
                               \hd_{l^2+2k+1}\ A_j^*\ \hd_{l^2+2k+1}\ \ ,
\\ \nonumber
    \Tr(A_j)\id_j +A_j\ =\ & 
               \frac{2}{d_j} \Big[\ \sum_{k=0}^{d_j-2}\sum_{l=k+1}^{d_j-1} 
               \hd_{l^2+2k}\ A_j^*\ \hd_{l^2+2k}\ +   \Big.
\\ \nonumber
       & +\ 
                \Big.\ \sum_{k=0}^{d_j-1} \hd_{k^2+k}\ A_j^*\ \hd_{k^2+k}\ \Big]
\\
                            \ =\ & 
               \frac{2}{d_j} \sum_{k=0}^{d_j-1}\sum_{l=k}^{d_j-1} 
               \hd_{l^2+2k}\ A_j^*\ \hd_{l^2+2k}   
\label{Seq:chan2}
\ \ .
\end{align}
As before, $A_j$ denotes a Hermitian operator acting on a $d_j$-dimensional 
Hilbert space.

The generalized $T$-inversion map for the state $\rho$ of an $N$-partite 
system has $N$ factors
as in the preceding equalities; the sign in front of the identity 
in the $j$th factor depends
on the presence of the index $j$ of the respective subsystem in $T$:
\begin{align}
   \mathcal{I}_T(\rho)\ =\ & \Big[ \prod_{j=1}^N 
                             \Big(
                           \Tr_j(\cdot)\otimes \id_j
                           +(-1)^{|T\cap \{j\}|}\mathrm{id}  
                             \Big)
                             \Big]\ \rho
\ , 
\label{Seq:T-inverter}
\end{align}
where $\mathrm{id}$ represents the identity map.
In order to rewrite Eq.~\eqref{Seq:T-inverter} we introduce the compact
notation
\begin{align*}
  k &= (k_1,\ldots,k_N)   \\
  t &= (t_1,\ldots,t_N)\ \ ,\ \ t_j=\frac{1}{2}\left[1-(-1)^{|\{j\}\cap T|}
                                           \right]
   \\
  \hd_{l^2+2k+t} &=
  \hd_{l_1^2+2k_1+t_1}\otimes\hd_{l_2^2+2k_2+t_2}\otimes\cdots\otimes
  \hd_{l_N^2+2k_N+t_N}   \nonumber\\
  \sum_{kl} &\equiv \sum_{k_1l_1} \cdots
  \sum_{k_Nl_N} \ \ , 
\end{align*}
where the $j$th tensor factor in $\hd_{l^2+2k+t}$
acts only on the $j$th subsystem and
the index ranges in the summation $\sum_{k_jl_j}$ need to be 
chosen as in Eq.~\eqref{Seq:chan1} if $j\in T$, or as in Eq.~\eqref{Seq:chan2}
otherwise. Then we can write 
\begin{align}
  \mathcal{I}_T(\rho)\ =\ \frac{2^N}{\prod_{j=1}^N d_j} \sum_{kl}
                          \hd_{l^2+2k+t}\ \rho^*\ \hd_{l^2+2k+t}
\ \ ,
\label{Seq:Kraus}
\end{align}
which is the desired Kraus form of the generalized $T$ inverter.

From Eq.~\eqref{Seq:Kraus} we readily see the product property
of the inverter on product states $\rho_{\mathrm{prod}} = 
\rho_S\otimes\rho_{S^c}$,
which is at the origin of Eq.~(19) in the main text:
\begin{align}
    \mathcal{I}_T\left(\rho_S\otimes\rho_{S^c}\right)\ =\ 
        \mathcal{I}_{T_S}(\rho_S)\otimes\mathcal{I}_{T_{S^c}}(\rho_{S^c})\ \ .
\label{Seq:prodprop}
\end{align}
Of course, this property is also evident from the product
representation of generalized $T$-inversion, Eq.~(7) in the main
text, and Eq.~\eqref{Seq:T-inverter} above.

\section{Coarse graining of the generalized $T$-inversion map}
\setcounter{equation}{0}
If we have a multipartite system with $N$ local parties, we can
choose to `coarse grain' the state by combining some of the 
parties, say $n$, into a single system.  We will show now how
the generalized $T$-inverter can be assembled from the inverters
of the `fine-grained' system. 
In principle, the $N$ local systems can be combined to
$k$ coarse-grained parties, where $1<k<N$. Due to the product structure of 
the generalized $T$-inversion map it is evident that it suffices
to understand how several parties can be combined to a {\em single}
party; the more general case of several coarse-grained parties
is obtained by applying the rules found for single-party coarse graining
to each combined party separately.
We denote the coarse-graining of
the multi-party into a single-party state by
\[
     \rho_{1\ldots n}\ \ \longrightarrow\ \ \rho_{(1\ldots n)}
\]
and the inverter on the combined system 
$\mathcal{I}_{T^{(1)}}(\rho_{(1\ldots n)})$, as opposed to that
of the fine-grained system, 
$\mathcal{I}_{T}(\rho_{1\ldots n})$.
Clearly, $T^{(1)}$ can equal $\emptyset$ or $\{1\}$,
corresponding to the two possible signs of the single-system inverter.
We will show now that
\begin{align}
    \mathcal{I}_{T^{(1)}}(\rho_{(1\ldots n)})\ =\
     \frac{1}{2^{n-1}}\sum_{S\subseteq \{1\ldots n\},\atop
              |S|\equiv |T^{(1)}|\!\!\!\!\! \pmod{2}}\!\!
                            \mathcal{I}_{S}(\rho_{1\ldots n})\ ,
\label{Seq:coarse}
\end{align}
which means, in order to obtain the coarse-grained inversion one has
to add all the fine-grained inversion operators whose parity coincides
with that of the desired coarse-grained operator. For example,
a minus inversion on a coarse-grained three-party system,
where \mbox{$T^{(1)}=\{1\}$}, is obtained via
\begin{align*}
   \mathcal{I}_{T^{(1)}}(\rho_{(123)})  \ =\ &
     \frac{1}{4}\big[
   \mathcal{I}_{\{1\}}(\rho_{123})\ +\
   \mathcal{I}_{\{2\}}(\rho_{123})\ + \big.
\\   & \big. \ \ \ \ \
   \mathcal{I}_{\{3\}}(\rho_{123})\ +\
   \mathcal{I}_{\{123\}}(\rho_{123})
                \big]\ \ .
\end{align*}
The proof is by induction. The case $n=1$ is trivial, as 
$\frac{1}{2^0}\sum^{\prime}_{S\subseteq \{1\}} \mathcal{I}_S = 
    \mathcal{I}_{T^{(1)}}$,
because the sum over $S$ contains only two terms $\emptyset$ and 
$\{1\}$, and we take into account (denoted by the prime)
only the term $|S|\equiv |T^{(1)}| \pmod{2}$.
Now we assume correctness of Eq.~\eqref{Seq:coarse} for $n$ parties and find for 
$(n+1)$-party coarse graining
\begin{widetext}
\begin{align}
  \mathcal{I}_{T^{(1)}}\left[\rho_{\big( 1\ldots (n+1)\big)}
                       \right]\ =\ &
  \Tr\left[\rho_{\big( 1\ldots (n+1)\big)}\right]\id 
                       + (-1)^{\left|T^{(1)}\right|}
       \rho_{\big( 1\ldots (n+1)\big)}
\nonumber\\
 =\ &  \frac{1}{2}\left( 
  \Tr\left[\rho_{\big( 1\ldots (n+1)\big)}\right]\id 
              + (-1)^{\left|T^{(1)}\right|} 
                \rho_{( 1\ldots n)}\otimes \id_{(n+1)}\ +\ 
              \id_{1\ldots n}\otimes \rho_{(n+1)}\ +\
                (-1)^{\left|T^{(1)}\right|} 
       \rho_{\big( 1\ldots (n+1)\big)}\ +\ \right.
\nonumber\\
    & \left.  +\
  \Tr\left[\rho_{\big( 1\ldots (n+1)\big)}\right]\id 
              - (-1)^{\left|T^{(1)}\right|} 
                \rho_{( 1\ldots n)}\otimes \id_{(n+1)}\ -\ 
              \id_{1\ldots n}\otimes \rho_{(n+1)}\ +\
                         (-1)^{\left|T^{(1)}\right|}
       \rho_{\big( 1\ldots (n+1)\big)}
                  \right)
\nonumber\\
 =\ &  \frac{1}{2}\Big( 
                        \mathcal{I}_{T^{(1)}}(\cdot)
                         \otimes
                       \left[\Tr_{(n+1)}(\cdot)\otimes\id_{(n+1)}+  
                                                 \mathrm{id}\right]
                        \rho_{(1\ldots n)(n+1)}   \Big.
           \ \ +
\nonumber\\      &  \Big. \ \ \ \ \ \ \ \ \ \ \ \ \ \ \ \ \ \ \ \ +\
                      \mathcal{I}_{\{1\}\setminus T^{(1)}}(\cdot)
                         \otimes
                       \left[\Tr_{(n+1)}(\cdot)\otimes\id_{(n+1)}-
                                                 \mathrm{id}\right]
                          \rho_{(1\ldots n)(n+1)}
                  \Big)
\ \ ,
\end{align}
\end{widetext}
where we can now make use of Eq.~\eqref{Seq:coarse}
\begin{widetext}
\begin{align}
  \mathcal{I}_{T^{(1)}}\left[\rho_{\big( 1\ldots (n+1)\big)}
                       \right]\ =\ &
          \frac{1}{2}\left( 
     \frac{1}{2^{n-1}}\sum_{S\subseteq \{1\ldots n\},\atop
              |S|\equiv |T^{(1)}|\!\!\!\!\! \pmod{2}}\!\!
                           \mathcal{I}_{S\cap\emptyset}(\rho_{1\ldots (n+1)})
\ +\
     \frac{1}{2^{n-1}}\sum_{S\subseteq \{1\ldots n\},\atop
              |S|\equiv |T^{(1)}|+1\!\!\!\!\! \pmod{2}}\!\!
                       \mathcal{I}_{S\cap\{(n+1)\}}(\rho_{1\ldots (n+1)})
                     \right)
\nonumber\\
 =\ &  
          \frac{1}{2^n}
               \sum_{S\subseteq \{1\ldots (n+1)\},\atop
              |S|\equiv |T^{(1)}|\!\!\!\!\! \pmod{2}}\!\!
                           \mathcal{I}_{S}(\rho_{1\ldots (n+1)})\ \ ,
\end{align}
\end{widetext}
which concludes the proof for $(n+1)$.
%

\section{Functional independence of the correlation constraints}
\setcounter{equation}{0}
In order to prove the functional indepedence for the
right-hand sides of the constraints in Eq.~(17) in the main text, 
we need to show that if for all~$\rho$,
\begin{equation}
  \mathscr D(\rho)\ \equiv\ \sum_{T \subseteq \{1\ldots N\} } 
                      \alpha_{T}
  \mathscr C_{T}^2(\rho) = 0
\label{Seq:Dsum}
\end{equation}
then $\alpha_{T}=0$ for all $T$.

We demonstrate this by constructing a family of states $\rho(S)$,
so that $\alpha_T=0$ is necessary in order to fulfill 
Eq.~\eqref{Seq:Dsum}.
Consider for all subsets $S \subseteq \{1\ldots N\}$ the
state $\rho(S) = \bigotimes_{k=1}^N \rho_k$ where
\begin{equation}
            \rho_k = \begin{cases}
                        \ket{0}\!\bra{0} & \text{ for } k \in S\\
                        \frac12(\ket{0}\!\bra{0}+\ket{1}\!\bra{1}) & \text{otherwise}\ \ .
                     \end{cases}
\label{Seq:rhok}
\end{equation}
A straightforward calculation gives
\begin{equation}
  \mathscr{C}_{T}^2(\rho(S)) = 
            \begin{cases}
         0 & \text{if } S \cap T \neq \emptyset\\
         \displaystyle
         \frac{4^{\left|S\right|}\cdot 
         3^{N-\left|S\right|-\left|T\right|}}{2^N} 
           & \text{if } S \cap T = \emptyset \ \ .
  \end{cases}
\end{equation}
Now we have $\rho(\{1\ldots N\})=\bigotimes_k \ket{0}\!\bra{0}$,
so that $\mathscr D\big(\rho(\{1\ldots N\})\big)=
\alpha_{\emptyset} \cdot 2^N = 0$, and hence $\alpha_{\emptyset}=0$.
Next we consider $S=\{1\ldots (k-1), (k+1)\ldots N\}$.
The only non-zero invariants for this $\rho(S)$ are $\mathscr{C}_{\emptyset}$
and $\mathscr{C}_{\{k\}}$. But we have already found that 
$\alpha_{\emptyset}=0$, hence also $\alpha_{\{k\}}=0$. By recursively
applying the same reasoning we conclude that $\alpha_S=0$ for
all $S\subseteq \{1\ldots N\}$.

Thus, we have proven independence of the $\mathscr{C}_T^2(\rho)$ for
mixed states. However, since we explicitly make statements also for
pure states (see Eq.~(21) in the main text), it is desirable to show
independence also for pure states. For this purpose, some preliminary 
observations are helpful.
First, we note that we may split the summation in Eq.~\eqref{Seq:Dsum}
\begin{align}
     \sum_{T\subseteq\{1\ldots N\}}\ =\
     \sum_{T^{\prime}=\emptyset,\{1\}}\ \ 
     \sum_{T^{\prime\prime}\subseteq\{2\ldots N\}}\ \ ,
\end{align}
and $T=T^{\prime}\cup T^{\prime\prime}$. Further, we have
from Eq.~\eqref{Seq:T-inverter}
\begin{align*}
   \mathcal{I}_{\emptyset\cup T^{\prime\prime}}(\psi)+
   \mathcal{I}_{\{1\}\cup T^{\prime\prime}}(\psi) &  =
   2\ \id_1\!\otimes\mathcal{I}_{T^{\prime\prime}}\left(
                                     \Tr_1\left[\ket{\psi}\!\bra{\psi}\right]                                                     \right)
 \ ,
\end{align*}
and therefore
\begin{align}
     \mathscr{C}^2_{\emptyset\cup T^{\prime\prime}}(\psi) +
     \mathscr{C}^2_{\{1\}\cup T^{\prime\prime}}(\psi)\ =\
   2 \mathscr{C}^2_{T^{\prime\prime}}\left(
                                     \Tr_1\left[\ket{\psi}\!\bra{\psi}\right]                                                     \right)
\  .
\label{Seq:Csum}
\end{align}
Moreover, on the left-hand side of Eq.~\eqref{Seq:Csum} only one of the terms
can be non-zero, because the other term has an odd number of minus signs
in the inverter (that is, $|T|=\left|T^{\prime}\cup T^{\prime\prime}\right|
\equiv 1\pmod{2}$) and therefore $\mathscr{C}_T^2(\psi)=0$, as we will
prove in Appendix D.

With the preceding remarks we conclude
\begin{align}
  \mathscr D(\psi) &\ \equiv\ \sum_{T \subseteq \{1\ldots N\} } 
                      \alpha_{T}
  \mathscr C_{T}^2(\psi)
\nonumber\\
     &\ =\ 2\sum_{T^{\prime\prime} \subseteq \{2\ldots N\}}
                      \alpha_{T^{\prime\prime}}
  \mathscr{C}_{T^{\prime\prime}}^2\left(
                                     \Tr_1\left[\ket{\psi}\!\bra{\psi}\right]                                                     \right)
\ \ ,
\end{align}
that is, we have reduced the $N$-qudit problem for pure states $\ket{\psi}$ to
an $(N-1)$-qudit problem for $\Tr_1\left(\ket{\psi}\!\bra{\psi}\right)$. 
Hence, in principle
we can use the proof for mixed states. The only remaining task is
to construct a family of pure states $\ket{\psi(S)}$ for which 
$\Tr_1\left(\ket{\psi(S)}\!\bra{\psi(S)}\right)$ has properties analogous to 
those of $\rho(S)$, see Eq.~\eqref{Seq:rhok}. 
Note that $S\subseteq \{2\ldots N\}$.  
An example for such a state is
\begin{align*}
   \ket{\psi(S)} = \oost\left[\ket{0}_1\bigotimes_{k\in S^c} \ket{0}_k
                              + \ket{1}_1\bigotimes_{l\in S^c} \ket{1}_l
                          \right]\otimes \bigotimes_{m \in S}
                                                    \ket{0}_m
\end{align*}
for which we find
\begin{equation}
  \mathscr{C}_{T}^2(\psi(S)) = 
            \begin{cases}
         0 & \text{if } S \cap T \neq \emptyset\\
         \displaystyle
         \delta_{0,|T|}2^{N-1}+2^{|S|} 
           & \text{if } S \cap T = \emptyset \  .
  \end{cases}
\end{equation}
With this, the proof can be completed as above for mixed states.

\section{Properties of the local unitary invariants $\mathscr{C}_T(\psi)$}
\setcounter{equation}{0}
First, let us prove Eq.~(19) in the main text.
To this end, consider for pure states $\rho=\ket{\psi}\!\bra{\psi}$
a term in the sum of Eq.~\eqref{Seq:Kraus}
when $|T|=m$ is an odd integer. Without loss of generality we
may assume that the minus sign occurs in the first $m$ parties, so that
\begin{align*}
 \hd_{l^2+2k+t} \ \rho^*\ \hd_{l^2+2k+t}\ =\ 
 \hd_{l^2+2k+t} \ \ket{\psi^*}\!\bra{\psi^*}\ \hd_{l^2+2k+t}
\end{align*}
and
\begin{align*}
 & \hd_{l^2  +2k+t}  \ = \ \\          
                & \ \ \ 
  \yd_{k_1l_1}\otimes\cdots \otimes\yd_{k_{m}l_{m}}\otimes 
          \hd_{l_{m+1}^2+2k_{m+1}}\otimes\cdots\otimes \hd_{l_N^2+2k_N}\ \ ,
\end{align*}
where the last $(N-m)$ tensor factors are of $\xd$ type or diagonal.
Those latter operators do not change under transposition or conjugation,
therefore, in what follows, we will not write them explicitly.

Now consider the corresponding term in the expansion of 
\begin{align*}
  \mathscr{C}_T^2(\psi)\ =\ \Tr\big(\ket{\psi}\!\bra{\psi}\ 
                               \mathcal{I}_T(\psi) \big)
\end{align*}
(for the sake of compactness we will drop also the 
symbol for the tensor product), we find
\begin{align*}
 \Tr & \big(\ket{\psi}\!\bra{\psi} \ 
         \hd_{l^2+2k+t} \ \ket{\psi^*}\!\bra{\psi^*}\ \hd_{l^2+2k+t}\big)
\ =\ \\
                &
 \bra{\psi}\yd_{k_1l_1}\cdots\yd_{k_{m}l_{m}}\cdots 
         \ \ket{\psi^*}\!\bra{\psi^*}\ 
 \yd_{k_1l_1}\cdots\yd_{k_{m}l_{m}}\cdots \ket{\psi}\ .
\end{align*}
That is, each term in the expansion of $\mathscr{C}_T^2(\psi)$ can be
written as
\begin{align*}
 \left|\bra{\psi}\yd_{k_1l_1}\cdots\yd_{k_{m}l_{m}}\cdots 
         \ \ket{\psi^*}\right|^2\ \ .
\end{align*}
But we have
\begin{align*}
 \bra{\psi}\yd_{k_1l_1}\cdots\yd_{k_{m}l_{m}}\cdots \ket{\psi^*} 
         &  = 
     \bra{\psi^*}\yd_{k_1l_1}^*\cdots\yd_{k_{m}l_{m}}^*\cdots \ket{\psi}^* 
\\
         &  = 
     -\bra{\psi}\yd_{k_1l_1}\cdots\yd_{k_{m}l_{m}}\cdots \ket{\psi^*} 
\end{align*}
for an odd number of $\yd$ type operators; therefore each term in
the expansion of $\mathscr{C}_T^2(\psi)$ vanishes, and 
\begin{align}   \mathscr{C}_T(\psi)\ =\ 0
\label{Seq:CT=0}
\end{align}
for odd $|T|=m$.

Let us finally prove that $\mathscr{C}_T(\psi)$ cannot be an entanglement
monotone if $T\neq \{1\ldots N\}$. 
First we remark that we need to consider only $T\neq \emptyset$ because
$\mathscr{C}_{\emptyset}^2(\psi)$ is a sum of local purities, and therefore
neither $\mathscr{C}_{\emptyset}(\psi)$ nor $\mathscr{C}_{\emptyset}^2(\psi)$
can be entanglement monotones (they are maximized on product states).
Similarly, a $\mathscr{C}_T(\psi)$ from a $T$-inverter with at least 
two plus signs in the product Eq.~\eqref{Seq:T-inverter} cannot be a 
monotone either. To see this, we note that the local invariant obeys the
product property on product states 
$\rho_{\mathrm{prod}}=\rho_{S}\otimes\rho_{S^c}$
\begin{align}
     \mathscr{C}_T(\rho_{\mathrm{prod}})\ =\ 
     \mathscr{C}_{T_S}(\rho_S) \mathscr{C}_{T_{S^c}}(\rho_{S^c})
\ \ ,
\end{align}
(which immediately follows from Eq.~\eqref{Seq:prodprop}).
Now consider a state $\ket{\psi_1}=\ket{\psi_S}\otimes\ket{\psi_{S^c}}$ 
such that $S^c$ contains all the parties with a plus sign in the inverter
(i.e., $T_S=T$). For a fully separable state $\ket{\psi_{S^c}}$, 
$\mathscr{C}_T(\psi_1)$ will then have a larger value than if we had chosen
an entangled state for $\psi_{S^c}$. Consequently, $\mathscr{C}_T(\psi)$
cannot be an entanglement monotone. The remaining case is that of 
a single plus sign in the inverter. There, we need to have at least {\em two}
minus signs [otherwise $\mathscr{C}_T(\psi)=0$ because of
Eq.~\eqref{Seq:CT=0}]. It is then easy to find counterexamples for the
monotone assumption. Consider, e.g., a system of three parties $(123)$ 
with local dimensions $d_j\geqq 2$,
$T=\{2,3\}$, and the state $\ket{\psi_{2}}=\oost(\ket{000}+\ket{111})$,
so that $\mathscr{C}_T(\psi_{\mathrm{I}})=1$. 
If we apply a two-outcome positive operator-valued measure (POVM) 
$\{A_1,A_2\}$ to the first qudit
with $A_{1,2}=\matel{\pm}{\pm}+\oost\sum_{j=2}^{d_1-1}\matel{j}{j}$,
$\ket{\pm}=\oost(\ket{0}\pm \ket{1})$, the resulting 
state is a tensor product of a pure state of the first party and a 
Bell-type state of the other two qudits,
so that the average of $\mathscr{C}_T$ for the two outcomes gives 
$\sqrt{2} > \mathscr{C}_T(\psi_{2})$, 
in contradiction with the monotone
assumption. Finally, for the case of a single plus sign and
a larger (even) number $k>2$ of minus signs in
the inverter we can construct an analogous counterexample 
$\ket{\psi_3}=\ket{\psi_2^{\prime}}\otimes \bigotimes_{l=1}^{\frac{k}{2}-1}
\oost\left(\ket{00}+\ket{11}\right)$;
here $\ket{\psi_2^{\prime}}$ is a state of three parties, where
the first party is the one with the plus sign in the inverter, in
analogy with $\ket{\psi_2}$.
The corresponding two-outcome POVM acts on that first party in
$\ket{\psi_2^{\prime}}$. Note that instead of
the tensor product of Bell-type states in $\ket{\psi_3}$ we could have
used any other state for which the inverter with only minus signs
does not vanish.
This concludes the proof that $\mathscr{C}_T(\psi)$ cannot be an
entanglement monotone for $T\neq\{1\ldots N\}$.

For completeness we mention that
for $T=\{1\ldots N\}$, $N$ even,  
the distributed concurrence $\mathscr{C}_T(\psi)$
is an entanglement monotone only in the following
cases (recall that for odd $N$ we have $\mathscr{C}_T\equiv 0$):\\
(a) $N=2$, $d_j$ arbitrary. Then $\mathscr{C}_T(\psi)$ coincides 
with the well-known concurrence for bipartite states.\\
(b) $d=2$, $N$ arbitrary. This case corresponds to the
polynomial invariant $|H(\psi)|=|\bra{\psi}\sigma_2^{\otimes N}\ket{\psi^*}|$,
which is the straightforward generalization of Wootters' two-qubit 
concurrence to $N$-qubit states.\\
(c) $N>2$, $d_j\leqq 3$. Also in those cases, $\mathscr{C}_T(\psi)$ 
is an entanglement monotone, as was proven in~\cite{ES2018}.

%
%
%

%
%

%



\begin{thebibliography}{99}
%
\bibitem{Bennett1996}
C.H.\ Bennett, D.P.\ DiVincenzo, J.A.\ Smolin, W.K.\ Wootters,
{\em Mixed-state entanglement and quantum error correction},
Phys.\ Rev.\ A {\bf 54}, 3824 (1996).
%
%
\bibitem{CKW2000} 
   V.\ Coffman, J.\ Kundu, and W.K.\ Wootters, 
{\em Distributed entanglement},
   Phys.\ Rev.\ A \textbf{61}, 052306 (2000).
%
\bibitem{Osborne2006}
T.J.\  Osborne and F.\ Verstraete, 
{\em General Monogamy Inequality for Bipartite Qubit Entanglement},
  Phys.\ Rev.\ Lett.\ {\bf 96}, 220503, (2006).
%
\bibitem{KW2004}
 M.\ Koashi and A.\ Winter, 
{\em Monogamy of entanglement and other correlations},
   Phys.\ Rev.\ A {\bf 69}, 022309 (2004).
%
\bibitem{Ou2007}
Y.C.\ Ou and H.\ Fan,
{\em Monogamy inequality in terms of negativity for three-qubit states}, 
Phys.\ Rev.\ A {\bf 75}, 062308 (2007).
%
\bibitem{Gour2007}
G.\ Gour,
{\em Dual monogamy inequality for entanglement},
J.~Math.\ Phys.\ {\bf 48}, 012108 (2007).
%
\bibitem{Kim2009}
J.S.\ Kim, A.\ Das, and B.C.\ Sanders,
{\em Entanglement monogamy of multipartite higher-dimensional 
     quantum systems using convex-roof extended negativity},
Phys.\ Rev.\ A {\bf 79}, 012329 (2009).
%
\bibitem{Eltschka2009}
C.\ Eltschka, A.\ Osterloh, and J.\ Siewert,
{\em Possibility of generalized monogamy relations for 
     multipartite entanglement beyond three qubits}
Phys.\ Rev.\ A {\bf 80}, 032313 (2009).
%
\bibitem{Sanders2010}
J.S.\ Kim and B.C.\ Sanders,
{\em Monogamy and polygamy for multi-qubit entanglement using R\'enyi entropy},
J.~Phys.\ A: Math.\ Theor.\ {\bf 43}, 442305 (2010).
%
\bibitem{Bai2014}
Y.-K.\ Bai, Y.-F.\ Xu, and Z.D.\ Wang,
{\em General Monogamy Relation for the Entanglement of Formation 
     in Multiqubit Systems},
Phys.\ Rev.\ Lett.\ {\bf 113}, 100503 (2014).
%
\bibitem{Fei2015}
X.-N.\ Zhu and S.-M.\ Fei,
{\em Generalized Monogamy Relations of Concurrence for $N$-qubit Systems},
Phys.\ Rev.\ A {\bf 92}, 062345 (2015).
%
\bibitem{ES2015}
  C.\ Eltschka and J.\ Siewert,
{\em Monogamy equalities for qubit entanglement from Lorentz invariance},
  Phys.\ Rev.\ Lett.\ {\bf 114}, 140402 (2015). 
%
\bibitem{Lancien2016}
C.\ Lancien, S.\ Di Martino, M.\ Huber, M.\ Piani, G.\ Adesso, and
A.\ Winter,
{\em Should Entanglement Measures be Monogamous or Faithful?}
Phys.\ Rev.\ Lett.\ {\bf 117}, 060501 (2016). 
%
\bibitem{Adesso2016}
B.\ Regula, A.\ Osterloh, and G.\ Adesso, 
{\em Strong monogamy inequalities for four qubits},
Phys.\ Rev.\ A {\bf 93}, 052338 (2016).
%
\bibitem{Luo2016}
Y.\ Luo, T.\ Tian, L.-H.\ Shao, and Y.\ Li,
{\em General Monogamy of Tsallis-q Entropy Entanglement in Multiqubit Systems},
Phys.\ Rev.\ A {\bf 93}, 062340 (2016).
%
\bibitem{Fei2017}
X.-N.\ Zhu, X.\ Li-Jost, and S.-M.\ Fei,
{\em Monogamy relations of concurrence for any dimensional quantum systems},
Quant.\ Inf.\ Process.\ {\bf 16}, 279 (2017).
%
\bibitem{Meyer2017}
G.W.\ Allen and D.A.\ Meyer,
{\em Polynomial Monogamy Relations for Entanglement Negativity},
Phys.\ Rev.\ Lett.\ {\bf 118}, 080402 (2017).
%
\bibitem{Gour2017}
G.\ Gour and Y.\ Guo,
{\em Monogamy of entanglement without inequalities},
e-print arXiv:1710.03295 (2017).
%
\bibitem{Huber2017}
P.\ Appel, M.\ Huber, and C.\ Kl\"ockl,
{\em Monogamy of correlations and entropy inequalities in 
the Bloch picture},
e-print arXiv:1710.02473 (2017).
%
\bibitem{ES2018}
  C.\ Eltschka and J.\ Siewert,
{\em Distribution of entanglement and correlations in all finite dimensions},
Quantum {\bf 2}, 64 (2018).
%
\bibitem{Toner2009}
B.\ Toner,
{\em Monogamy of nonlocal correlations}, 
Proc.\ R.\ Soc.\ A {\bf 465}, 59 (2009).
%
%
\bibitem{Seevinck2010}
M.P.\ Seevinck, 
{\em Monogamy of Correlations vs. Monogamy of Entanglement}
Quant.\ Inf.\  Proc.\ {\bf 9}, 273 (2010).
%
\bibitem{Cabello2017}
Z.-A.\ Jia, G.-D.\ Cai, Y.-C.\ Wu, G.-C.\ Guo, and A.\ Cabello,
{\em The Exclusivity Principle Determines the Correlation Monogamy},
e-print arXiv:1707.03250 (2017).
%
\bibitem{Ujjwal2016}
H.\ Shekhar Dhar, A.\ Kumar Pal, D.\ Rakshit, A.\ Sen De, 
and U.\ Sen,
{\em Monogamy of quantum correlations - a review},
e-print  arXiv:1610.01069 (2016).
%
\bibitem{Horodecki1999}
M.\ Horodecki and P.\ Horodecki,
{\em Reduction criterion of separability and limits for a class of distillation protocols},
Phys.\ Rev.\ A {\bf 59}, 4206 (1999).
%
%
\bibitem{Rungta2001}
P.\ Rungta, V.\ Buzek, C.M.\ Caves, M.\ Hillery, and G.J.\ Milburn,
{\em Universal state inversion and concurrence in arbitrary dimensions},
Phys.\ Rev.\ A {\bf 64}, 042315 (2001).
%
\bibitem{Rains1998}
E.M.\ Rains, 
{\em Quantum weight enumerators},
IEEE Trans.\ Inf.\ Theory {\bf 44}, 1388 (1998).
%
\bibitem{Rains2000}
E.M.\ Rains, 
{\em Polynomial invariants of quantum codes},
IEEE Trans.\ Inf.\ Theory {\bf 46}, 54 (2000).
%
\bibitem{Audenaert2007}
K.\ Audenaert,
{\em Subadditivity of $q$-entropies for $q>1$},
J.~Math.\ Phys.\ {\bf 48}, 083507 (2007).
%
\bibitem{Lieb1973}
E.H.\ Lieb and M.B.\ Ruskai, 
{\em Proof of the strong subadditivity of quantum-mechanical entropy},
J.~Math.\ Phys.\ {\bf 14}, 1938 (1973).
%
\bibitem{Hayden2013}
P.\ Hayden, M.\ Headrick, and A.\ Maloney,
{\em Holographic Mutual Information is Monogamous},
Phys.\ Rev.\ D {\bf 87}, 046003 (2013).
%
\bibitem{McGill1954}
W.J.\ McGill,
{\em Multivariate information transmission},
Psychometrika {\bf 19}, 97 (1954).
%
%
\bibitem{Hall2005}
W.\ Hall, 
{\em Multipartite reduction criteria for separability},
Phys.\ Rev.\ A {\bf 72}, 022311 (2005).
%
\bibitem{Hall2006}
W.\ Hall,
{\em A new criterion for indecomposability of positive maps},
J.\ Phys.\ A: Math.\ Gen.\ {\bf 39}, 14119 (2006).
%
\bibitem{GellMann}
We use the definitions 
$  \mathsf{x}_{kl} =
  \sqrt{\frac{d_j}{2}}(\matel{k}{l}+\matel{l}{k})$, \linebreak
$  \mathsf{y}_{kl} =
  \sqrt{\frac{d_j}{2}}(-\mathrm i\matel{k}{l}+\mathrm i\matel{l}{k})$
for the off-diagonal matrices, and 
$ \mathsf{z}_l  = \sqrt{\frac{d_j}{l(l+1)}}
  \left(-l\matel{l}{l} + \sum_{k=0}^{l-1}\matel{k}{k}\right)$ 
for the diagonal generators
%
($0\leqq k < l < d_j$). 
%
\bibitem{supplement} see Supplemental Material.
%
\bibitem{AHW2000}
G.G.\ Amosov, A.S.\ Holevo and R.F.\ Werner,
{\em On some additivity problems in Quantum Information Theory};
Problems in Information Transmission, vol.~36, 305–313 (2000)
%
\bibitem{footnote-shadow}
Note that the left-hand side of\ Eq.~\eqref{eq:positive}
leads to the coefficients of Rains' shadow enumerator 
$S_j(M_1,M_2)$ if all expressions of equal
$|T|=N-j$ are added: $S_j(M_1,M_2)=\sum_{|T|=N-j}\sum_S(-1)^{|S\cap T|}
                                   \Tr\big[\Tr_{S^c}(M_1)\Tr_{S^c}(M_2)
                                       \big]$,
see Refs.~\cite{Rains1998,Huber2018}.
%
\bibitem{Sloane1990}
J.H.\ Conway and N.J.A.\ Sloane,
{\em A new upper bound on the minimal distance of self-dual code},
IEEE Trans.\ Inf.\ Theory {\bf 36}, 1319 (1990).
%
\bibitem{Dougherty1995}
S.T.\ Dougherty,
{\em Shadow codes and weight enumerators},
IEEE Trans.\ Inf.\ Theory {\bf 41}, 762 (1995).
%
\bibitem{Rains1998b}
E.M.\ Rains,
{\em Shadow bounds for self-dual codes},
IEEE Trans.\ Inf.\ Theory {\bf 44}, 134 (1998).
%
\bibitem{Rains1999}
E.M.\ Rains,
{\em Quantum shadow enumerators},
IEEE Trans.\ Inf.\ Theory {\bf 45}, 2361 (1999).
%

\bibitem{Huber2018}
F.\ Huber, C.\ Eltschka, J.\ Siewert, and O.\ G\"uhne,
{\em Bounds on absolutely maximally entangled states
from shadow inequalities, and the quantum
MacWilliams identity},
J.~Phys.\ A: Math.\ Theor.\ {\bf 51}, 175301 (2018).
%
%
\bibitem{Breuer2006}
H.-P.\ Breuer,
{\em Optimal Entanglement Criterion for Mixed Quantum States},
Phys.\ Rev.\ Lett.\ {\bf 97}, 080501 (2006).
%
\bibitem{Wolf2016}
A.\ M\"uller-Hermes, D.\ Reeb, and M.M.\ Wolf,
{\em Positivity of linear maps under tensor powers},
J.~Math.\ Phys.\ {\bf 57}, 015202 (2016).
%
\bibitem{Lewenstein2016}
M.\ Lewenstein, R.\ Augusiak, D.\ Chru\'sci\'nski, S.\ Rana,
and J.\ Samsonowicz,
{\em Sufficient separability criteria and linear maps},
Phys.\ Rev.\ A {\bf 93}, 042335 (2016).
%
\bibitem{Huber2016}
L.\ Lami and M.\ Huber,
{\em Bipartite depolarizing maps},
J.~Math.\ Phys.\ {\bf 57}, 092201 (2016).
%
\bibitem{Tsallis1988}
C.\ Tsallis,
{\em Possible generalization of Boltzmann-Gibbs statistics},
J.\ Stat.\ Phys.\ {\bf 52}, 479 (1988).
%
\bibitem{Wehrl1978}
A.\ Wehrl,
{\em General properties of entropy},
Rev.\ Mod.\ Phys.\ {\bf 50}, 221 (1978).
%
\bibitem{Petz2015}
D.\ Petz and D.\ Virosztek, 
{\em Some inequalities for quantum Tsallis entropy related to the 
strong subadditivity},
Math.\ Inequal.\ Appl.\ {\bf 18}, 555 (2015).
%
\bibitem{Lieb1970}
H.\ Araki and E.H.\ Lieb,
{\em Entropy inequalities},
Comm.\ Math.\ Phys.\ {\bf 18}, 160 (1970).
%
\bibitem{Uhlmann1973}
A.\ Uhlmann,
{\em Endlich-dimensionale Dichtematrizen II}, 
Wiss.\ Z.\ Karl-Marx-Univ.\ Leipzig, Math.-Nat.\ R.\ {\bf 22}, 139 (1973).
%
\bibitem{Klyachko2006}
A.A.\  Klyachko, 
{\em Quantum marginal problem and N-representability},
J.\ Phys.: Conf.\ Ser.\ {\bf 36}, 72 (2006).
%
\bibitem{Butterley2006}
P.\ Butterley, A.\ Sudbery, and J.\ Szulc,
{\em Compatibility of Subsystem States},
Found.\ Phys.\ {\bf 36}, 83 (2006).
%
\bibitem{Huber2017Thesis}
F.\ Huber, {\em  Quantum states and their marginals: 
from multipartite entanglement to quantum error-correcting codes},
\href{http://dokumentix.ub.uni-siegen.de/opus/volltexte/2018/1272/}
{Ph.D.\ Thesis}, 
University of Siegen, 2017; available at
{\tt http://dokumentix.ub.uni-siegen.de/\allowbreak opus/\allowbreak volltexte/\allowbreak 2018/1272/}.
%
\end{thebibliography}
\end{document}